# Magnetic-field-induced robust zero Hall plateau state in MnBi$_2$Te$_4$ Chern insulator

Chang Liu[1,2,†], Yongchao Wang[3,†], Ming Yang[4,†], Jiahao Mao[1], Hao Li[5,6], Yaoxin Li[1], Jiaheng Li[1], Haipeng Zhu[4], Junfeng Wang[4], Liang Li[4], Yang Wu[6,7], Yong Xu[1,8,9*], Jinsong Zhang[1,9*], Yayu Wang[1,9*]

[1]*State Key Laboratory of Low Dimensional Quantum Physics, Department of Physics, Tsinghua University, Beijing 100084, P. R. China*

[2]*Beijing Academy of Quantum Information Sciences, Beijing 100193, P. R. China*

[3]*Beijing Innovation Center for Future Chips, Tsinghua University, Beijing 100084, P. R. China*

[4]*Wuhan National Magnetic Field Center, Huazhong University of Science and Technology, Wuhan 430074, P. R. China*

[5]*School of Materials Science and Engineering, Tsinghua University, Beijing, 100084, P. R. China*

[6]*Tsinghua-Foxconn Nanotechnology Research Center, Department of Physics, Tsinghua University, Beijing 100084, P. R. China*

[7]*Department of Mechanical Engineering, Tsinghua University, Beijing 100084, P. R. China*

[8]*RIKEN Center for Emergent Matter Science, Wako, Saitama 351-0198, Japan*

[9]*Frontier Science Center for Quantum Information, Beijing 100084, P. R. China*

[†] *These authors contributed equally to this work.*

* Emails: yongxu@tsinghua.edu.cn; jinsongzhang@tsinghua.edu.cn; yayuwang@tsinghua.edu.cn

**Abstract**

The intrinsic antiferromagnetic topological insulator $MnBi_2Te_4$ provides an ideal platform for exploring exotic topological quantum phenomena. Recently, the Chern insulator and axion insulator phases have been realized in few-layer $MnBi_2Te_4$ devices at low magnetic field regime. However, the fate of $MnBi_2Te_4$ in high magnetic field has never been explored in experiment. In this work, we report transport studies of exfoliated $MnBi_2Te_4$ flakes in pulsed magnetic fields up to 61.5 T. In the high-field limit, the Chern insulator phase with Chern number $C = -1$ evolves into a robust zero Hall resistance plateau state. Nonlocal transport measurements and theoretical calculations demonstrate that the charge transport in the zero Hall plateau state is conducted by two counter-propagating edge states that arise from the combined effects of Landau levels and large Zeeman effect in strong magnetic fields. Our result demonstrates the intricate interplay among intrinsic magnetic order, external magnetic field, and nontrivial band topology in $MnBi_2Te_4$.

**Introduction**

A remarkable breakthrough in the field of topological quantum matter is the discovery of topological insulators (TIs) with nontrivial bulk band topology and metallic boundary states[1-8]. For two-dimensional (2D) TI with time-reversal symmetry (TRS), the helical edge states give rise to the quantum spin Hall (QSH) effect[3,5,9-11]. When TRS is broken in magnetic TI, the quantum anomalous Hall (QAH) effect with chiral edge state emerges[6-8,12-14]. Because the QAH effect originates from topological Chern band rather than Landau levels[13,14], it is now called the Chern insulator phase to distinguish it from conventional quantum Hall (QH) insulator[2,15,16]. Previous efforts on the Chern insulator mainly focused on its realization in zero magnetic field[6-8,12]. An intriguing question that has yet to be addressed experimentally is the fate of Chern insulator in ultra-high magnetic fields[17,18]. It is likely that the QAH plateau will not survive forever because extreme conditions may induce other topological phases, as exemplified by the fractional QH effect[19]. Recently, it was demonstrated that helical phases analogous to the QSH phase exist in charge-neutral graphene[20,21] and non-symmorphic KHgSb crystal[22] in strong magnetic fields, which are characterized by quantized longitudinal resistance ($R_{xx}$) and zero

Hall resistance ($R_{yx}$) plateau in certain magnetic field and gate voltage ($V_g$) ranges. Inspired by these discoveries, it is imperative to find out the fate of the Chern insulator phase in MnBi$_2$Te$_4$ in strong magnetic fields, as has been discussed theoretically for magnetic TIs[23,24].

The recently discovered MnBi$_2$Te$_4$ combines intrinsic magnetism and nontrivial topology in one material[25-37], providing an ideal platform for exploring topological phenomenon in extreme physical conditions. Figure 1a displays the schematic magnetic and crystal structure of MnBi$_2$Te$_4$, where the Mn$^{2+}$ magnetic moments have ferromagnetic (FM) alignments within each septuple layer (SL) and antiferromagnetic (AFM) coupling between neighboring SLs. In 2D limit, it exhibits the QAH insulator[30] and axion insulator[29] phases for odd- and even-number of SLs. When FM order is induced by magnetic fields, the parity-time (PT) symmetry is broken and the bulk of MnBi$_2$Te$_4$ becomes a Weyl semimetal[26,34,38]. In thin MnBi$_2$Te$_4$ flakes, the bulk Weyl semimetal band develops 2D quantum well states due to the quantum confinement along the $c$-axis. The topologically nontrivial quantum well subbands lead to robust Chern insulator state with quantized Hall plateau that persists to relatively high temperatures for magnetic field $\mu_0 H > 8$ T[29-31]. A unique feature of the Chern insulator phase in MnBi$_2$Te$_4$ lies in the negative sign of the Chern number ($C = -1$) in positive magnetization, which is distinguishable from the $C = +1$ Chern insulator phase in Cr- or V-doped TIs [6-8,12]. Figure 1b illustrates the spatial configurations of the chiral edge state and the band structures for the two cases. Phenomenologically, the opposite sign of Chern number arises from opposite effective magnetic field due to the interplay between magnetic order and spin-orbit coupling[39,40]. Because magnetic field couples with both magnetic moment and electron spin, for certain spin configurations of band structure, the combination of Landau levels and a sufficiently strong Zeeman energy may lead to topological quantum phenomenon such as TRS-broken QSH effect or the quantum pseudospin Hall effect[23,24] which have been proposed theoretically but never been realized in experiment.

In this work, we report transport studies on exfoliated MnBi$_2$Te$_4$ in pulsed magnetic fields up to 61.5 T. Unexpectedly, the $C = -1$ phase evolves into a zero Hall plateau state characterized by a broad $R_{yx} = 0$ plateau and insulating $R_{xx}$ in ultrahigh magnetic field. Nonlocal transport measurements and theoretical calculations demonstrate the transport of this zero Hall plateau state is composed of two counter-propagating edge states with opposite Chern numbers, which arise from the FM order and the joint roles of Landau levels and Zeeman effect. The robust zero Hall plateau state represents a topological phenomena that is unavailable in 2D electron- or hole-gas with conventional QH effect.

## Results

The MnBi$_2$Te$_4$ devices studied in this work are mechanically exfoliated few-layer flakes fabricated into field-effect transistor devices on SiO$_2$/Si substrates that act as the bottom gate. The details of fabrication and transport measurements in pulsed magnetic fields are described in the method session. All the data in the main text are collected on a 7-SL device (#7-SL-1), and its photo is displayed in Fig. 1c. We first characterize the low-field transport properties at temperature $T = 2$ K in varied $V_g$s, and three representative curves are shown in Fig. 1d to 1f (see Supplementary Fig. 1 for the full data set). The insets schematically illustrate the Fermi level ($E_F$) position for each $V_g$. The most pronounced feature is that for $V_g = 4$ V when $E_F$ is in the Chern band gap at FM state. At low magnetic field side, $R_{xx}$ exhibits three jumps due to the successive flips of Mn$^{2+}$ moments, which is consistent with previous reports on few-layer MnBi$_2$Te$_4$ devices[29-31]. Meanwhile, $|R_{yx}|$ progressively grows with magnetic field, and the slope is mainly determined by the magnetization. A well-defined quantum plateau forms at -$h/e^2$ in $R_{yx}$ for $\mu_0 H > 8$ T, accompanied by the rapid decrease of $R_{xx}$ to zero. These are consistent with the Chern insulator behaviors in previously reports[29-31]. As $V_g$ is tuned away from 4 V to either side, $E_F$ moves out of the band gap and carriers from the 2D subbands appear. The quantization of Chern insulator phase is suppressed, and the negative (positive) slopes of the Hall traces at $V_g$ = +16 V (-16 V) indicate the existence of electron (hole) type carriers. According to the linear slope, the mobility in the electron- and hole-type regime is estimated to be 3114 and 2098 cm$^2$/Vs respectively.

When we extend the transport measurements to much higher magnetic fields, some totally unexpected features start to emerge. According to the characteristic behavior of $R_{yx}$, the entire $V_g$ range can be divided into four different regimes. As shown in Fig. 2a, the most remarkable feature of the high-field data is that the $C = -1$ state only survives in a range of about 10 T. With further increase of magnetic fields, $R_{yx}$ drops rapidly from the $-h/e^2$ plateau and, even more surprisingly, a very broad $R_{yx} = 0$ plateau forms in the high-field regime. Meanwhile, $R_{xx}$ takes off from the zero value, develops a shoulder ~ 0.5 $h/e^2$ at the onset field of the zero plateau, and then increases again in higher magnetic fields. Remarkably, all the $R_{xx}$ curves for varied $V_g$s tend to converge to the 0.5 $h/e^2$ value (marked by the broken line) near the onset magnetic field of the zero Hall plateau. With the decrease of $V_g$, the high field zero Hall plateau is universally present and becomes even broader. As shown in Fig. 2b for $V_g$ from -2 V and 0 V, the zero Hall plateau spans an incredibly wide field range from 10 T to the highest available field of 61.5 T. In the Supplementary Fig. 2, we display the magnetic field dependent $R_{xx}$ and $R_{yx}$ at $V_g = 4$ V for varied temperatures, in which a clear tendency of $R_{xx}$ saturation at 0.5 $h/e^2$ is observed at low temperatures. Both the $C = -1$ phase and the zero Hall plateau state exhibit high robustness against thermal activation, even up to $T = 20$ K. Similar transport behaviors are also observed in other samples with different size and thickness (#7-SL-2 and #6-SL-1), as shown the Supplementary Fig. 3 to Fig. 7.

When $V_g$ is decreased to more negative values, as shown in Fig. 2c, the Hall traces evolve to that characteristic of a 2D hole gas with an overall positive profile and growing amplitude from $V_g = -4$ V to -16 V. Meanwhile, $R_{xx}$ reduces systematically as more holes are injected into the MnBi$_2$Te$_4$ flake. At $V_g = -16$ V, $R_{yx}$ forms a well-defined Hall plateau at $h/e^2$ ($C = +1$) for $\mu_0 H > 40$ T, whereas $R_{xx}$ drops to zero within experimental uncertainty. When $V_g$ is tuned to the opposite side with $V_g$ from 8 V to 16 V (Fig. 2d), quantized $R_{yx}$ plateaus with Chern numbers $C = -3, -2$ and $-1$ show up, along with apparent quantum oscillations. Remarkably, a unique feature here is that all the plateaus exhibit a strong tendency towards the zero Hall plateau state in the highest magnetic field, regardless of the position of $E_F$, which is quite unusual in conventional QH effect[16].

Based on the high field data shown above, in Figs. 3a and 3b we depict the contour maps of $R_{yx}$ and $R_{xx}$ in the $\mu_0 H$ and $V_g$ plane. The most prominent feature is that the zero Hall plateau state occupies the largest portion of the phase diagram. The magenta line marks the boundary between the $C = -1$ and zero Hall plateau state. The biggest puzzles revealed by the high field experiments are the nature of the zero Hall plateau state that prevails in the phase diagram and the underlying mechanism for the phase transition. In recent years, the zero Hall plateau states discovered in graphene and TI have attracted intense attentions[22,41-43] for their nontrivial origins. However, most previous works focused on Hall conductivity ($\sigma_{xy}$) rather than Hall resistivity ($\rho_{yx}$). Because $\sigma_{xy} = \rho_{yx}/(\rho_{xx}^2+\rho_{yx}^2)$, any kind of insulating state with large longitudinal resistivity $\rho_{xx}$ can give rise to a $\sigma_{xy} = 0$ plateau. In contrast, the observation of $R_{yx} = 0$ plateau is very rare in experiment. Moreover, a zero Hall plateau state evolved from a Chern insulator or QH state is highly unusual.

The zero Hall plateau and tendency of $R_{xx}$ towards 0.5 $h/e^2$ at the onset regime are highly reminiscent of the QSH effect, where the transport is conducted by a pair of helical edge states[3]. Differently, the helical transport in QSH insulator is protected by TRS[3,9,10], i.e., the absence of both magnetic field and magnetic order. But here, the zero Hall plateau state is observed when both FM order in strong magnetic field are present. Therefore, the zero Hall plateau state here represents a distinct insulator state that may host counter-propagating edge states transport with tunable scatterings in magnetic fields. To clarify this issue, we first calculate the band structure of 7-SL MnBi$_2$Te$_4$ when it is the FM state, as shown in Fig. 3c. According to theories, the breaking of PT-symmetry by FM order drives the bulk of MnBi$_2$Te$_4$ from an AFM TI into a Weyl semimetal[26,34,38]. In the thin film case, the low-energy physics near $E_F$ is described by four quantum well states, in analogy to the four-band model for the QAH effect in magnetic TIs[13]. In these bands, one pair of subbands (red curves) is already inverted due to FM order, which is responsible for the $C = -1$ phase. The blue bands represent two trivial quantum-well states with an energy gap as small as 3 meV.

The most intuitive explanation for the magnetic-field-induced zero Hall plateau state is to consider the coexistence of a hole-type Landau level ($C = +1$) and the FM-order induced Chern band ($C = -1$). However, because Landau level spacing due to cyclotron motion increases in magnetic field, the $C = -1$ phase for $E_F$ lying in the band gap will persist in strong magnetic field. Apparently, this is inconsistent with the experimental observation of the $C = -1$ to zero

Hall plateau transition. Therefore, it is indispensable to consider the band shift under Zeeman effect. Because a small Zeeman energy insufficient for band gap closing will not change the $C$ = -1 phase in certain $V_g$ range with $E_F$ located in the band gap, a sufficiently large Zeeman effect that can cause a band inversion must be included to explain the absence of persistent $C$ = -1 regime and the emergent zero Hall plateau state. The detailed illustrations of the band structure evolutions are displayed in the Supplementary Fig. 8.

A closer examination of the spin configuration of each band reveals an enticing possibility of the Zeeman-effect induced band inversion, as shown in Fig. 3d. Because of the strong spin-orbit coupling in MnBi$_2$Te$_4$, the $z$ component of spin angular momentum $s_z$ is no longer a good quantum number. Therefore, we label each state at the Γ point by the total angular momentum $J_z$, which is quantized by $C_3$ rotational symmetry and is mainly contributed by $s_z$ because of the small orbital angular momentum of $p_z$ orbitals. The black arrows in the band structure denote the $z$ component of $J_z$ for each band at the Γ point. An important point is that spin up refers to positive $J_z$ or equivalently negative spin magnetic moment $M_z$. Thus, in an external magnetic field, along with the formation of Landau levels, bands with opposite $J_z$ shift towards opposite direction. When $E_F$ initially lies near the top of the valence band (e.g. $V_g$ = 0 V), the formation of $n$ = 0 Landau level gives rise to a $C$ = +1 chiral edge state. In combination of the original FM-order induced $C$ = -1 band, a zero Hall plateau forms. Meanwhile, the Zeeman effect pushes the $n$ = 0 Landau level upwards in magnetic field, which leads to the $C$ = -1 to zero Hall plateau transition for $E_F$ lying the band gap (e.g. $V_g$ = 4 V). As magnetic field is increased further, the Zeeman energy can surpass the gap size of the trivial bands, and a band inversion happens at some critical field. Meanwhile, at the gap closing point, the $n$ = 0 Landau level crosses from the valence band to the inverted conduction band. Therefore, the total Chern number of the zero plateau state before and after band inversion does not change. The different shadows in Fig. 3d denote the regimes with different Chern numbers. This scenario is further supported by the calculated Landau level spectrums with the Zeeman effect taken into consideration, as highlighted by the upward phase boundary (magenta line) in Fig. 3e. The In the Supplementary Fig. 10, we also calculate the Landau level spectrum without Zeeman effect, which displays apparently qualitative departure from our experimental phase diagram.

For a better comparison with the previously discovered helical QH phase in graphene[20,21], we zoom in the $R_{xx}$ and $R_{yx}$ data in the Chern insulator regime, as displayed in Fig. 4a. At $V_g$ = 2 V, the width of the $R_{xx}$ ~ 0.5 $h/e^2$ plateau is as broad as 10 T. As $V_g$ is increased to 4 V, the plateau becomes a broad shoulder. In the inset of Fig. 4a, we display the schematic illustrations of measurement setups and the evolution of counter-propagating edge states in magnetic field. The opposite chirality of the two edge channels at the onset of the zero Hall plateau can naturally explain the zero Hall plateau and the quantized behaviors in $R_{xx}$. When the new $C$ = +1 edge state just appears, its spatial distribution is not confined to the sample boundary[44]. The scattering between the two counter-propagating edge states is weak, thus the $R_{xx}$ value at the initial stage of the zero Hall plateau state is close to 0.5 $h/e^2$ for scattering-immune helical transport, which is exactly the scenario for stabilizing the QSH edge state with broken TRS[44]. With the further increase of magnetic field, the edge states are pushed towards the boundaries. Because there is no TRS, the enhanced overlap between the two edge states leads to strong scatterings, giving rise to a rapid increase of $R_{xx}$. Similar situations of the deviations of $R_{xx}$ in high magnetic field are also observed in the helical QH phase in graphene[20,21].

To further validate the counter-propagating edge state nature of the zero Hall plateau state, we perform multiterminal and nonlocal transport measurements at $V_g$ = 4 V, as displayed in Fig. 4b. The schematic setups of the measurements are shown in the insets. In the situation without scatterings between two edge states, the Landauer-Buttiker formalism renders $R_{2T}$ = 2 $h/e^2$ and $R_{3T}$ = 1/2 $h/e^2$. Because each pair of neighboring electrodes contributes a quantum resistance, the final resistance can be calculated by an effective circuit determined by the arrangement of electrodes between source and drain[3,4]. The measured results exactly match such expectation, as marked by the magenta dashed lines. The middle and bottom panels show the results of two different setups for nonlocal measurements, which can directly detect edge state transport[4]. The measured 1/8 $h/e^2$, 3/4 $h/e^2$ and 1/4 $h/e^2$ are the expected values for counter-propagating edge transport with weak scattering. At high magnetic field where scatterings are enhanced, the zero Hall plateau state evolves towards a trivial insulator when the counter-propagating edge states are fully canceled. However, the nonlocal transport data shows that even up to 61.5 T, there is still a sizeable edge state conduction. For the setup in the middle panel of Fig. 4b, at the onset of the zero Hall plateau when the scattering is weak, the ratio between local and nonlocal resistance is 1:4, which is fully determined by the electrode configurations. The ratio remains at 1:4 at high magnetic field, which is a characteristic signature of edge state transport. Because a true trivial insulator has no edge conductivity, and residual carriers cannot give rise to such robust zero Hall plateau and do not contribute to any nonlocal signal, the zero Hall plateau state observed here is a distinct insulator phase from a true trivial insulator. Reproducible non-local data with much broader plateau width can be seen in another 7-SL sample (#7-SL-3) shown in Supplementary Fig. 11.

## Discussion

The physical picture of emergent Chern band gaps induced by Landau levels and Zeeman effect gives a comprehensive understanding of the zero Hall plateau state observed here in thin flakes of MnBi$_2$Te$_4$ in strong magnetic fields. The key factor is to create a Chern band gap in magnetic field so that a new $C = +1$ edge state can be involved in transport. The Landau levels guarantee the formation of Chern band gaps inside the conduction or valence bands, whereas the Zeeman effect ensures the inversion of the original trivial bands in magnetic field. Only in this case the universal tendency towards a zero Hall plateau state throughout the Chern insulator and the QH regimes can be well explained.

Naively, a zero Hall resistance in magnetic fields can be also attributed to other origin such as the coexistence of electron-hole puddles in magnetic field. But a careful analysis of the transport data can rule it out. First, the zero Hall resistance due to the exact cancellation between electron and hole puddles is an accidental state that cannot form a broad $R_{yx} = 0$ plateau in a wide range of magnetic field and $V_g$. It cannot explain the convergence of $R_{xx}$ towards the 0.5 $h/e^2$ plateau in the initial stage of zero Hall plateau state either. In addition, a trivial insulator is not a likely explanation either because it will exhibit diverging $R_{yx}$ in strong magnetic field in dc-transport measurement, rather than a broad zero Hall plateau[45,46]. The presence of edge transport throughout the magnetic field regime, as well as the low-temperature saturation of $R_{xx}$ for different magnetic field, also distinguish the zero Hall plateau state from a trivial insulator, as shown in Supplementary Fig. 2b. Notably, it is plausible that a spin polarized edge mode with ballistic transport length beyond our longest channel length ~ 10.8 μm emerges in high magnetic field. It may also lead to plateau-like feature with quantized $R_{xx}$. Further studies are required to completely rule out this possibility.

In conclusion, we realize a zero Hall plateau state in MnBi$_2$Te$_4$ Chern insulator state when a strong magnetic field is applied. The robust zero Hall plateau against magnetic field and $V_g$ as well as the nonlocal transport measurements suggest this state is composed of two counter-propagating edge states that arise from the emergence of a new Chern band gap. The zero Hall plateau state discovered in magnetic field in MnBi$_2$Te$_4$ represents a unique quantum transport phenomenon generated by the intricate interplay among intrinsic magnetism, external magnetic field and nontrivial band topology.

**Methods**

**Crystal growth** High-quality MnBi$_2$Te$_4$ single crystals were grown by direct mixture of Bi$_2$Te$_3$ and MnTe with the ratio of 1:1 in a vacuum-sealed silica ampoule. After first heated to 973 K, the mixture is slowly cooled down to 864 K, followed by a long period of annealing process. The phase and crystal quality are examined by X-ray diffraction on a PANalytical Empyrean diffractometer with Cu Kα radiation.

**Device fabrication** MnBi$_2$Te$_4$ flakes were mechanically exfoliated onto 285 nm-thick SiO$_2$/Si substrates by using the Scotch tape method. Before exfoliation, all SiO$_2$/Si substrates were pre-cleaned in air plasma for 5 minutes with ~ 125 Pa pressure. Thick flakes around the target sample area were manually scratched off by using a sharp needle. A 270 nm thick Poly(methyl methacrylate) (PMMA) layer was spin-coated on the exfoliated film before electron-beam lithography (EBL). After the EBL, 53 nm thick metal electrodes (Cr/Au, 3/50 nm) were deposited using a thermal evaporator connected to an argon-filled glove box with the O$_2$ and H$_2$O levels lower than 0.1 PPM. Throughout the fabrication and sample transfer process, the device was covered by PMMA to avoid direct contact with air. Four devices with 7-SL and 6-SL MnBi$_2$Te$_4$, denoted as device #7-SL-1, #7-SL-2, #7-SL-3 and #6-SL-1, were measured in pulsed magnetic field. The data shown in the main figures are taken from device #7-SL-1, and that of other devices are documented in Supplementary Fig. 1 to Fig.11.

**Transport measurement** High-field electrical transport measurements were performed in a $^4$He cryostat with the base temperature of 2 K in Wuhan National High Magnetic Field Center. A pulsed DC current of 4 µA was generated by a Yokogawa GS610 current source. An uncertainty of 250 Ω arising from the high rate of field sweep (1000 T/s) in pulsed magnetic field measurements and imperfect cancellation of measurement circuit was estimated according to the real geometry of the circuit. The absence of hysteresis in low-field transport data of the 7-SL MnBi$_2$Te$_4$ samples is due to fast field sweeping rate. Low-field calibration of the 6-SL thick device was performed in a commercial $^4$He cryostat with a superconducting magnet up to 9 T. The longitudinal and Hall voltages were measured simultaneously by using lock-in amplifiers with AC current of 200 nA generated by a Keithley 6221 current source. The back gate was applied by a Keithley 2400 source meter. To eliminate the effect of electrode misalignment, the measured four-terminal longitudinal and transverse resistances were symmetrized and antisymmetrized with respect to magnetic field.

**Theoretical calculation** First-principles calculations were performed in the framework of density functional theory (DFT) using the Vienna *ab initio* Simulation Package[47]. The plane-wave basis with an energy cutoff of 350eV was adopted, in combination with the projected augmented wave (PAW) method. The Monkhorst-Pack *k*-point mesh of $9 \times 9 \times 3$ were adopted in the self-consistent calculation with inclusion of spin-orbit coupling. The modified Becke-Johnson (mBJ) functional[48] was employed to improve the description of electronic band structure in the ferromagnetic (FM) MnBi$_2$Te$_4$ bulk. The DFT-D3 method[49] was used to describe van der Waals (vdW) interactions between neighboring septuple layers in MnBi$_2$Te$_4$. The tight-binding models derived from the FM bulk were used to model thin films. Maximally localized Wannier functions were constructed from first-principles calculations of the FM bulk, and the tight-binding Hamiltonian of the FM bulk was obtained. Then, tight-binding Hamiltonians of thin films were constructed by cutting slabs from the bulk. Band structures, topological properties, edge state calculations[50] and effective $\boldsymbol{k}\cdot\boldsymbol{p}$ Hamiltonians of MnBi$_2$Te$_4$ thin films were computed based on the tight-binding Hamiltonians.

**Data Availability:** All raw and derived data used to support the findings of this work are available from the authors on request.

**Acknowledgements:** This work is supported by the Basic Science Center Project of NSFC (grant No. 51788104) and National Key R&D Program of China (grants No. 2018YFA0307100 and No. 2018YFA0305603). This work is supported in part by the Beijing Academy of Quantum Information Sciences (BAQIS) and Beijing Advanced Innovation Center for Future Chip (ICFC).

**Competing interests:** The authors declare no competing interests.

**Author contributions:** Y. Y. W., J. S. Z. and Y. X. supervised the research. C. L., Y. C. W. and Y. X. L. fabricated the devices and performed the transport measurements. M. Y., H. P. Z., J. F. W. and L. L. were in charge of the pulsed magnet facility. H. L. and Y. W. grew the MnBi$_2$Te$_4$ crystals. J. H. M, J. H. L. and Y. X. performed first-principles calculations. C. L., J. S. Z, Y. X. and Y. Y. W. prepared the manuscript with comments from all authors.

**FIGURE CAPTIONS**

**Figure 1 | Basic properties of a 7-SL MnBi$_2$Te$_4$. a**, Schematic crystal and magnetic structures of the 7-SL MnBi$_2$Te$_4$ device. **b,** Configurations of chiral edge state in the Chern insulator with Chern number $C$ = -1 and +1. The opposite chirality of edge state is marked by red and blue lines with arrows. The magenta arrows denote the magnetic moments. The schematic electronic structures for the two cases are shown on the right, with the opposite slope of the linear band representing opposite chirality. **c**, Optical image of Device #7-SL-1 and the measurement setup. **d,** Magnetic field dependent $R_{xx}$ (red) and $R_{yx}$ (blue) at $V_g$ = 4 V and $T$ = 2 K. The Chern insulator phase is realized when magnetic field is above 8 T, which is characterized by the $R_{yx}$ = -$h/e^2$ plateau and $R_{xx}$ = 0. **e**, As $V_g$ is tuned to -16 V, the transport is dominated by hole-type carriers. The jumps in $R_{xx}$ at magnetic field of around 1.8 T, 4 T, and 7 T correspond to the successive flips of Mn$^{2+}$ moments in different SLs. **f**, At $V_g$ = 16 V, the $E_F$ is tuned to the conduction band and the transport exhibits characteristic features of 2D electron gas. The insets in **d** to **f** roughly show the position of $E_F$ at according $V_g$.

**Figure 2 | Transport properties in pulsed magnetic field up to 61.5 T. a,** Magnetic field dependent $R_{xx}$ and $R_{yx}$ at 1 V $\leq V_g \leq$ 6 V. At $V_g$ = 4 V, the $C$ = -1 state is completely suppressed for $\mu_0 H$ > 30 T, followed by the $C$ = 0 state characterized by a broad zero Hall plateau. The black dashed line denotes the $R_{xx}$ = 0.5 $h/e^2$ plateau. **b**, Transport properties at -2 V $\leq V_g \leq$ 0 V. Zero Hall plateau exist in a broad magnetic field range over 50 T. **c**, Transport behaviors in the 2D hole gas regime. With the decrease of $V_g$, QH plateaus with positive Chern number start to form. At $V_g$ = -16 V, $R_{yx}$ forms a wide Hall plateau at $h/e^2$ and $R_{xx}$ drops to zero. **d**, Characteristic transport behaviors of 2D electron gas. With the increase of $V_g$ from 8 V to 16 V, the $C$ = -1 Hall plateau onsets at a higher magnetic field, becomes broader, and approaches the $C$ = 0 plateau only at the high-field limit. Electron-type QH plateaus with higher Chern number $C$ = -2 and -3 also start to form.

**Figure 3 | Contour plots of experimental data and theoretical analysis of the $C$ = -1 to $C$ = 0 phase transition. a-b,** Experimental phase diagrams of $R_{yx}$ and $R_{xx}$ in the magnetic field and

$V_g$ plane. The $C = 0$ phase is the most stable ground state in strong magnetic field. The magenta broken line denotes the boundary between the $C = -1$ and $C = 0$ phase. The black arrow represents the regime of $R_{xx} \sim 0.5\ h/e^2$ for the helical Chern insulator phase. **c,** Calculated band structure of 7-SL MnBi$_2$Te$_4$ along the M-Γ-K direction when the system is driven into the FM state. The red and blue lines denote the ferromagnetic-order induced Chern band ($C = -1$) and topologically trivial band ($C = 0$) respectively. **d,** Schematic illustrations of the edge state formation and the band structure evolution in magnetic field with Zeeman-effect-induced band inversion. The black and red dashed lines roughly mark the $E_F$ position for $V_g = 4$ V and 2 V. The $C = -1$ to $C = 0$ phase transition occurs once band inversion happens. **e,** Calculated Landau level spectrums with Zeeman-effect.

**Figure 4 | Signatures of helical edge states transport in multiterminal and nonlocal measurements in the $C = 0$ phase. a,** Magnetic field dependent $R_{xx}$ and $R_{yx}$ near the $C = -1$ to $C = 0$ phase transition for $V_g = 2$ V and 4 V. The spatial distribution of edge states in magnetic field is displayed in the inset. **b,** Two-, three-terminal and nonlocal measurements in various configurations. The inset shows the schematic layout of the experimental setup. The expected values for $R_{2T}$ and $R_{3T}$ are 2 $h/e^2$ and 1/2 $h/e^2$, as denoted by the broken lines in the top panels. Middle panel: nonlocal measurements with current flowing through electrodes 1 and 8. The convergence of all three curves near 1/8 $h/e^2$ (denoted by the broken lines) indicates the helical edge transport at the onset of the $C = 0$ phase. Bottom panel: nonlocal measurements in another setup with current flowing through electrodes 1 and 7. Depending on the position of the voltage probes, the resistance values of 1/4 and 3/4 $h/e^2$ are expected, which are confirmed by the experimental results.

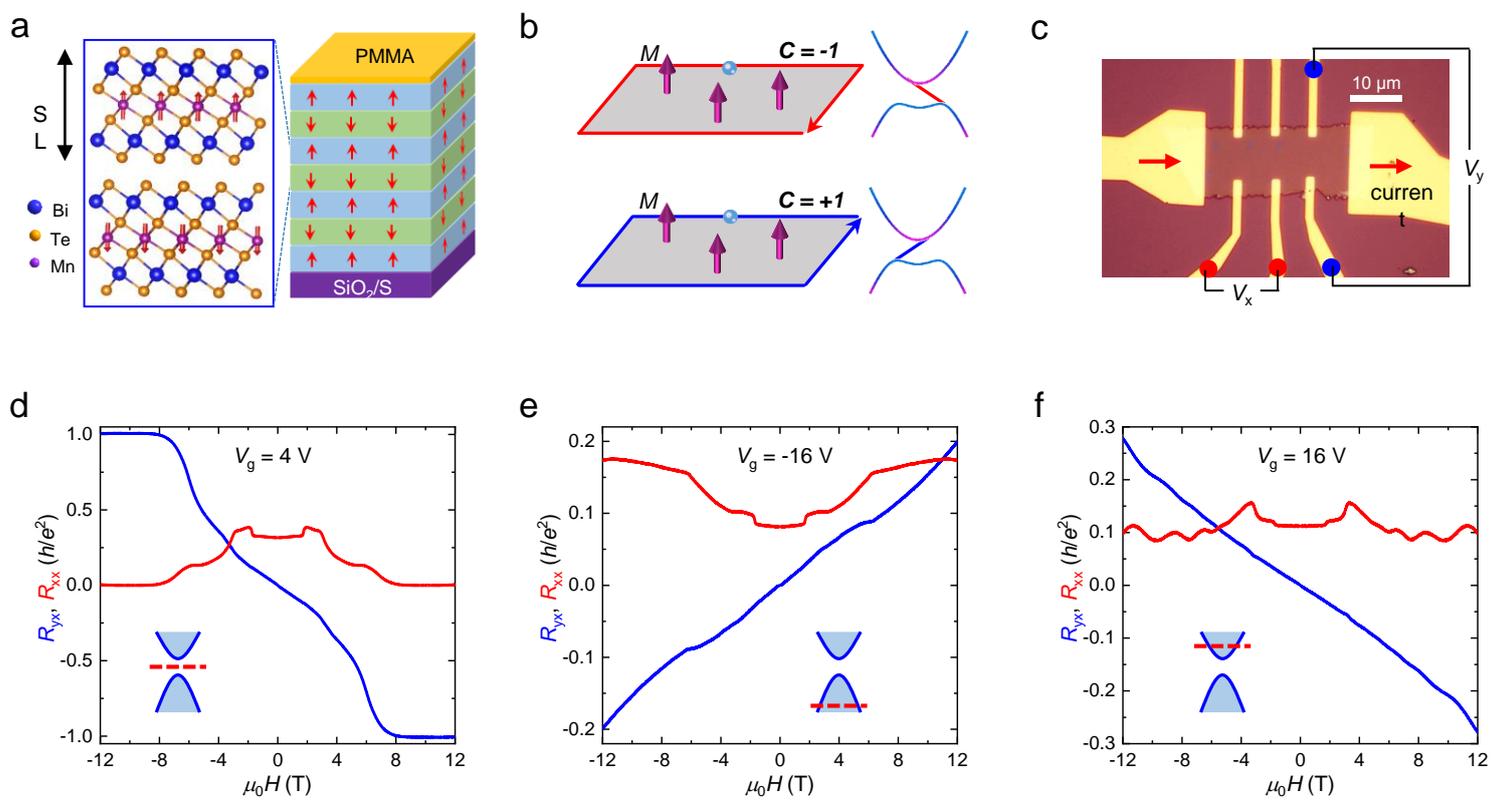

Figure 1

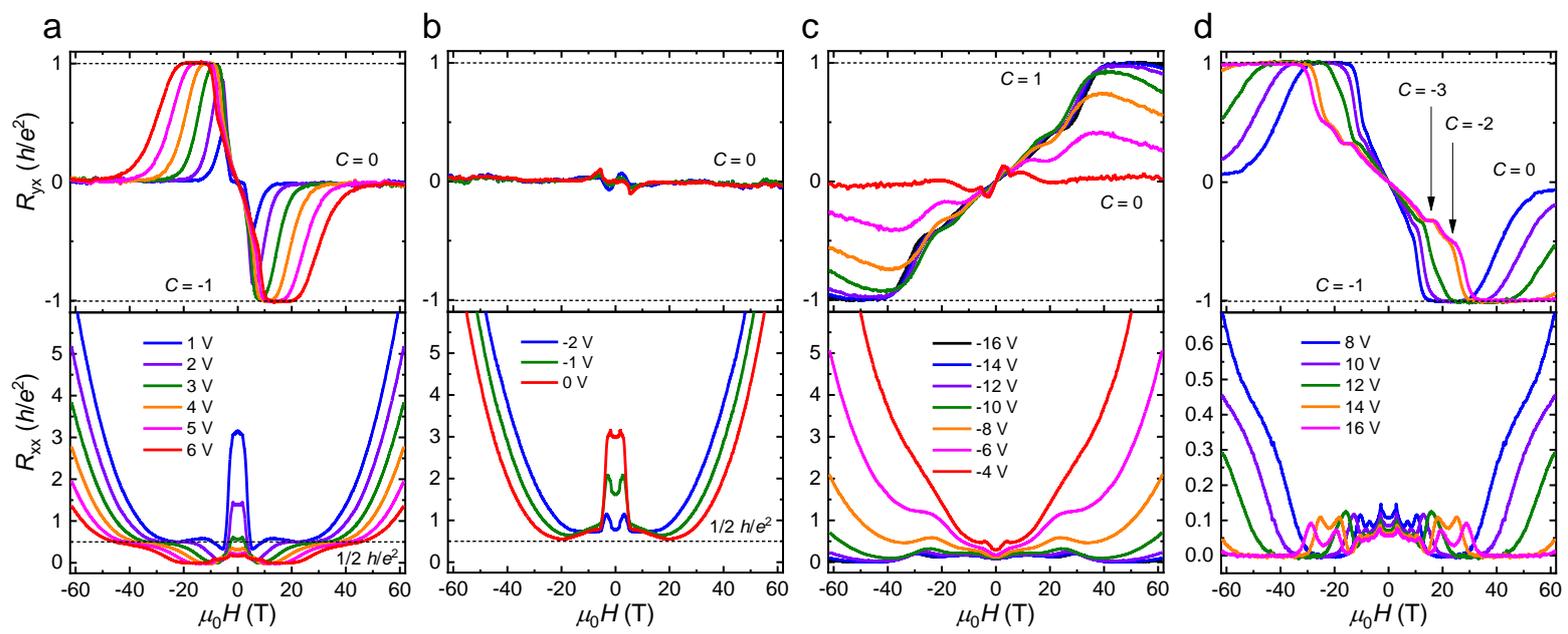

Figure 2

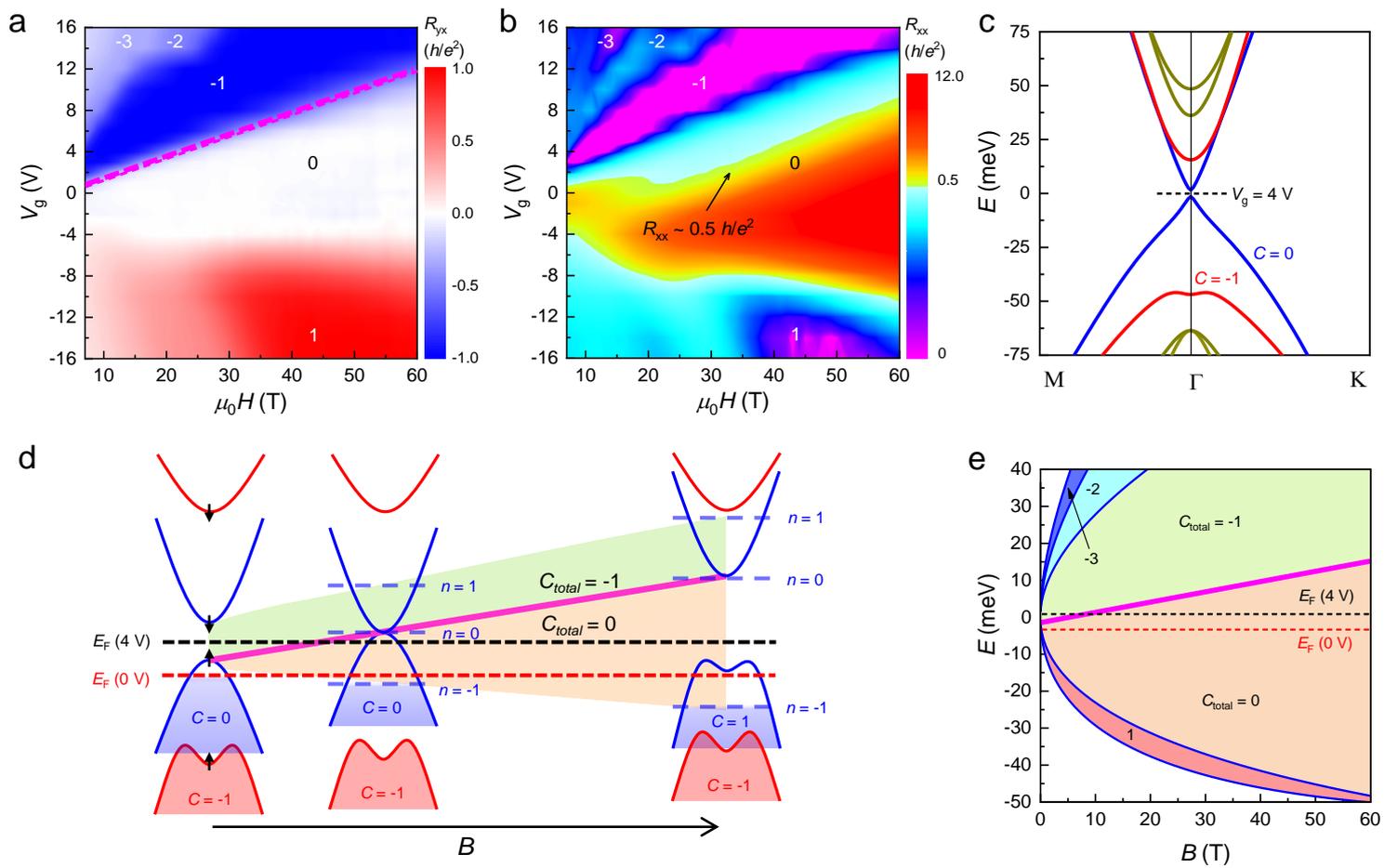

Figure 3

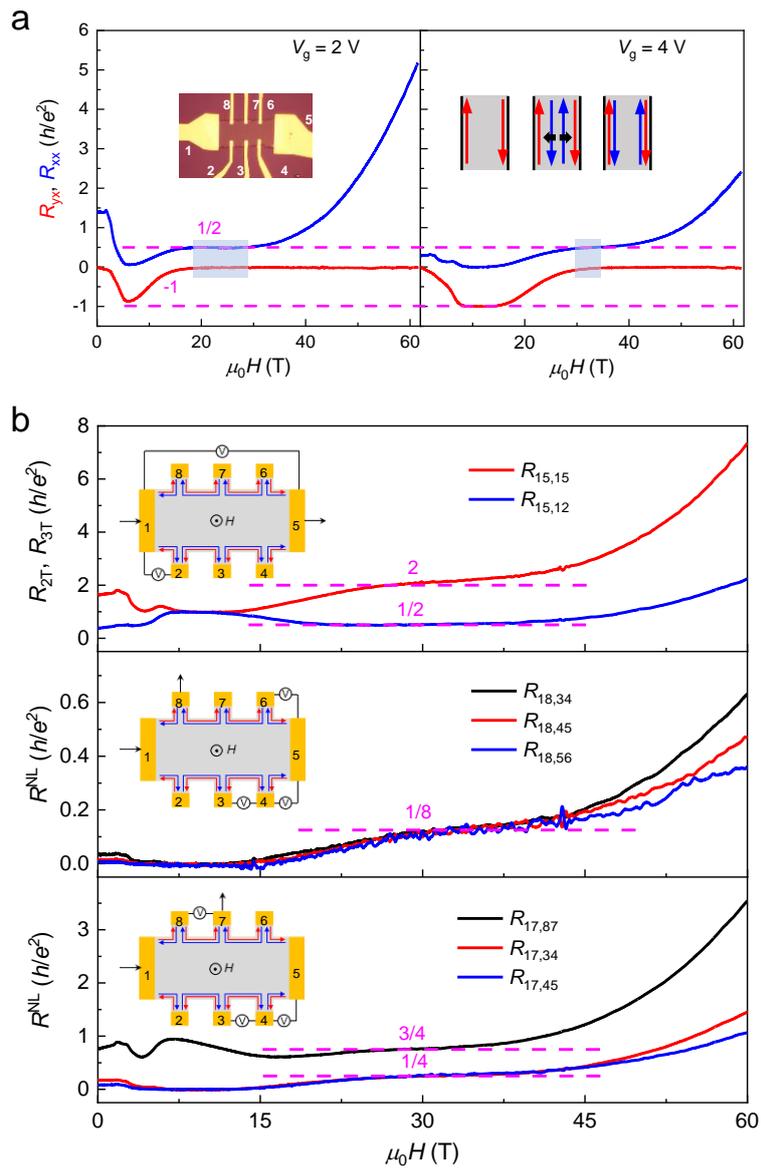

Figure 4

# Supplementary Information

# Magnetic-field-induced robust zero Hall plateau state in MnBi$_2$Te$_4$ Chern insulator

Chang Liu[1,2,†], Yongchao Wang[3,†], Ming Yang[4,†], Jiahao Mao[1], Hao Li[5,6], Yaoxin Li[1], Jiaheng Li[1], Haipeng Zhu[4], Junfeng Wang[4], Liang Li[4], Yang Wu[6,7], Yong Xu[1,8,9*], Jinsong Zhang[1,9*], Yayu Wang[1,9*]

[1]*State Key Laboratory of Low Dimensional Quantum Physics, Department of Physics, Tsinghua University, Beijing 100084, P. R. China*

[2]*Beijing Academy of Quantum Information Sciences, Beijing 100193, P. R. China*

[3]*Beijing Innovation Center for Future Chips, Tsinghua University, Beijing 100084, P. R. China*

[4]*Wuhan National Magnetic Field Center, Huazhong University of Science and Technology, Wuhan 430074, P. R. China*

[5]*School of Materials Science and Engineering, Tsinghua University, Beijing, 100084, P. R. China*

[6]*Tsinghua-Foxconn Nanotechnology Research Center, Department of Physics, Tsinghua University, Beijing 100084, P. R. China*

[7]*Department of Mechanical Engineering, Tsinghua University, Beijing 100084, P. R. China*

[8]*RIKEN Center for Emergent Matter Science, Wako, Saitama 351-0198, Japan*

[9]*Collaborative Innovation Center of Quantum Matter, Beijing, P. R. China*

[†] These authors contributed equally to this work.

* Emails: yongxu@tsinghua.edu.cn; jinsongzhang@tsinghua.edu.cn; yayuwang@tsinghua.edu.cn

**Supplementary Note:**

**1. Low-field transport properties of Device #7-SL-1**

**2. Temperature dependent transport behavior for Device #7-SL-1**

**3. Pulsed field transport data for Device #7-SL-2**

**4. Basic low-field transport calibrations for Device #6-SL-1**

**5. $V_g$ dependent $R_{xx}$ and $R_{yx}$ for Device #6-SL-1 in pulsed magnetic fields**

**6. Experimental phase diagram of the 6-SL device**

**7. Schematic band structure evolution for different situations of Zeeman effect**

**8. Effective Hamiltonian and calculated Landau levels**

**9. Reproducible nonlocal transport data for Device #7-SL-3**

**Supplementary Note 1:**

**Low-field transport properties of Device #7-SL-1**

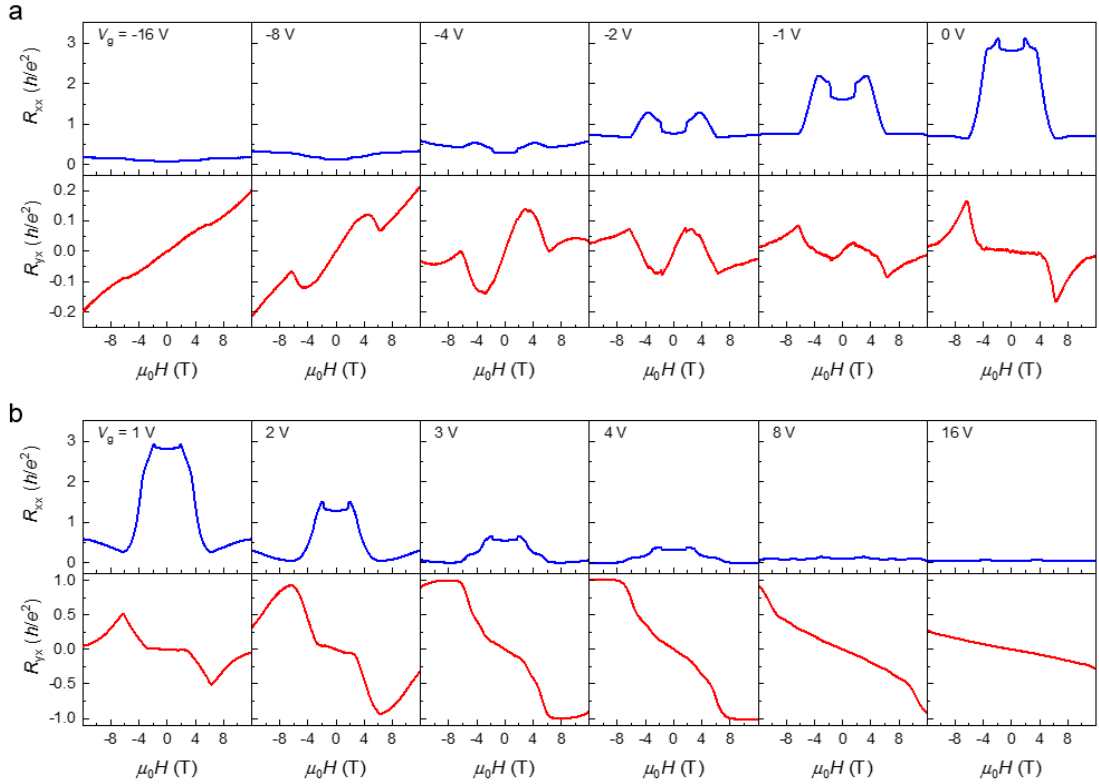

**Supplementary Fig. 1 | Low-field transport properties of the 7-SL MnBi$_2$Te$_4$ device S1. Magnetic field dependent $R_{xx}$ and $R_{yx}$ measured at 2 K for -16 V ≤ $V_g$ ≤ 0 V (a), and 1 V ≤ $V_g$ ≤ 16 V (b).**

In this work, we investigated the transport properties of four MnBi$_2$Te$_4$ samples with different thickness and sample size, including three 7-SL devices (#7-SL-1, #7-SL-2 and #7-SL-3) and one 6-SL device (#6-SL-1). The results presented in the main figures were taken from Device #7-SL-1. Due to the limited space, only part of the data are displayed.

In this session, we present the complete data set of the magnetic field dependent $R_{xx}$ and $R_{yx}$ at varied $V_g$s, as shown in Supplementary Figs. 1a and 1b, respectively. In the regime of $V_g \leq 0$ V, the low-field $R_{xx}$ increases with $V_g$, and the low-field slope of $R_{yx}$ changes sign at $V_g = 0$ V. With the increase of magnetic field, a series of jumps appear in both $R_{xx}$ and $R_{yx}$, corresponding to the sequential flipping events of magnetic moments in each individual SL of $MnBi_2Te_4$. Both $R_{xx}$ and $R_{yx}$ show abrupt changes at magnetic field of 6 T, indicating the dramatic change of band structure upon entering the FM state. At $V_g = 4$ V, the system enters the Chern insulator phase when FM order forms, as discussed in the main text. Further increase of $V_g$ raises the position of $E_F$ towards the conduction band, leading to more conductive properties. In addition to the decrease of $R_{xx}$, $|R_{yx}|$ also deviates from the quantized plateau ($h/e^2$) and develops a negative slope that is characteristic of ordinary Hall effect in a 2D electron gas.

**Supplementary Note 2:**

**Temperature dependent transport behavior for Device #7-SL-1**

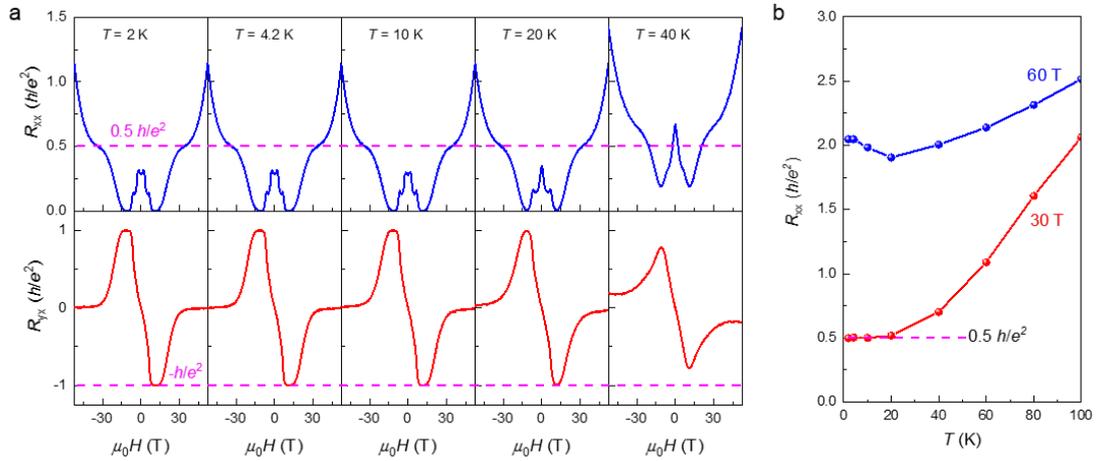

**Supplementary Fig. 2 | Temperature dependent transport behavior for Device #7-SL-1 at $V_g = 4$ V. a**, Magnetic field dependent $R_{xx}$ and $R_{yx}$ measured at 2 K. **b**, Temperature dependent $R_{xx}$ in different magnetic fields.

Supplementary Fig. 2a shows the magnetic field dependent $R_{xx}$ and $R_{yx}$ for Device #7-SL-1 at $V_g = 4$ V and varied temperatures from $T = 2$ K to 40 K. Both the $C = -1$ and $C = 0$ phases are highly robust against temperature. Even at $T = 20$ K, the quantization of $R_{yx}$ reaches as high as $-0.991\ h/e^2$ for the $C = -1$ phase. The zero Hall plateau for the helical $C = 0$ phase and the quantization of $R_{xx}$ at $0.5\ h/e^2$ near 30 T are also very clear. As the temperature is increased further, significant changes appear in $R_{xx}$ at low magnetic field regime, which is related to the weakening of AFM order above the Neel temperature $T_N \sim 25$ K. Meanwhile, both the features of the $C = -1$ and $C = 0$ phases are weakened. These results unambiguously show the robustness of the Chern insulator phase in high magnetic field. In Supplementary Fig. 2b, we display the temperature dependent $R_{xx}$ for the $C = 0$ phase in different magnetic fields. At the onset field of about 30 T, the scattering between the helical edge states is weak, so the half-quantized $R_{xx}$ persists in a broad temperature regime of 20 K, as marked by the magenta dashed line. At 60 T where the scattering leads to more insulating behavior, the value of $R_{xx}$ is much larger. However, fundamentally different from trivial insulator with diverging $R_{xx}$ with lowering temperature, $R_{xx}$ increases slightly and saturates at the ground state. The temperature dependent $R_{xx}$ clearly shows that the helical $C = 0$ phase in our work is not a trivial insulator.

## Supplementary Note 3:
## Pulsed field transport data for Device #7-SL-2

To further demonstrate the reproducibility of the main results presented in the main text, we measured another 7-SL device (#7-SL-2) in pulsed magnetic fields. The optical image of this device is shown in Supplementary Fig. 3a. During the cooling process, one current lead had bad connection to electrode 1 in the image, therefore the current was applied from electrode 6 to 4. The measurement setup for $R_{xx}$ and $R_{yx}$ are marked in the figure. Supplementary Fig. 3b displays the magnetic field dependent $R_{xx}$ and $R_{yx}$ measured at different $V_g$s. The overall behaviors of both $R_{xx}$ and $R_{yx}$ are consistent to that observed in Device #7-SL-1 (see Fig. 2 in the main figures for details), although there are some quantitative differences in the detailed magnetic field and gate voltage dependences.

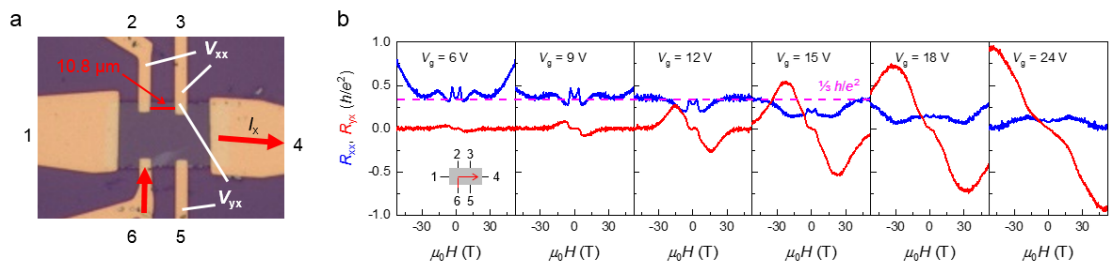

**Supplementary Fig. 3 | Magnetic field dependent $R_{xx}$ and $R_{yx}$ measured in pulsed**

fields for Device #7-SL-2 at various $V_g$s. a, Optical images and measurement setups for Device #7-SL-2. b, Magnetic field dependent $R_{xx}$ and $R_{yx}$ for different $V_g$s. The magenta dashed lines mark the quantized values of ⅓ $h/e^2$ expected for $R_{xx}$ at the onset of zero Hall plateau.

The most striking observation is that a broad zero Hall plateau also forms in the high field regime in Device #7-SL-2. With the increase of $V_g$, the zero Hall plateau becomes narrower, and the $C$ = -1 plateau starts to emerge. Notably, a broad quantized $R_{xx}$ ~ ⅓ $h/e^2$ plateau appears at the onset magnetic field of the $C$ = 0 phase, as marked by the magenta broken line. This is exactly the quantized resistance expected for helical edge states in such electrode configuration.

Another important point is that Device #7-SL-2 has larger size than Device #7-SL-1 shown in the main text. The distance between the two $R_{xx}$ electrodes is 10.8 μm, whereas for Device #7-SL-1 it is only 6.7 μm. The quantization of $R_{xx}$ at the onset of the zero Hall plateau in samples with larger size is another strong evidence to support the helical nature of the $C$ = 0 phase.

**Supplementary Note 4:**
**Basic low-field transport calibrations for Device #6-SL-1**

In this session, we present the magnetic field dependent transport results measured at different temperatures and $V_g$s for an even-number-layer sample (Device #6-SL-1). The variation of film thickness mainly affects the low-field properties when $MnBi_2Te_4$ is in the AFM state, which is not the focus of this work. In the FM state, both 6-SL and 7-SL samples are expected to exhibit the Chern insulator behavior with robust quantized Hall plateau[1,2].

Supplementary Fig. 4 shows the magnetic field dependent $R_{xx}$ and $R_{yx}$ for the 6-SL device at varied $V_g$s. A unique feature for 6-SL $MnBi_2Te_4$ device is that the axion insulator phase is expected at low-field regime when $E_F$ is tuned to the CNP, as demonstrated in our previous report[1]. The overall behaviors of this device is similar to that of the three 7-SL devices. Both $R_{xx}$ and $R_{yx}$ show systematic evolutions in response to magnetic field. At the magnetic-field-driven AFM to FM transition, $R_{xx}$ exhibits a dramatic decrease, accompanied by an abrupt jump in $R_{yx}$. At $V_g < 36$ V, the low-field $R_{xx}$ increases with the increase of $V_g$, and $R_{yx}$ exhibits linear behavior with overall positive slope in magnetic field. At 42 V $\leq V_g \leq$ 46 V, the system enters the regime of axion and Chern insulator phases. In the low-field regime for the axion insulator phase, the value of $R_{xx}$ reaches as high as 5 $h/e^2$, followed by a sharp decrease to zero as magnetic field exceeds 6 T. Accordingly, $R_{yx}$ exhibits a sharp transition from the zero plateau to the $R_{yx} = -h/e^2$ plateau. As $V_g$ is further increased, the system deviates from the axion and Chern insulator phases, as a natural result of the increased contributions from electron-type charge carriers. At $V_g = 59$ V, $R_{xx}$ is reduced to as low as 1 $h/e^2$, and $R_{yx}$ exhibits overall negative slope in both the low- and high-field regime. Noting that larger $V_g$ is required in the 6-SL device for observing the Chern insulator phase, which suggests that the 6-SL device is much more hole-doped than the 7-SL device. This is because fabrication process tends to introduce holes in the pristine electron-doped bulk crystal, as is shown in our previous report[1].

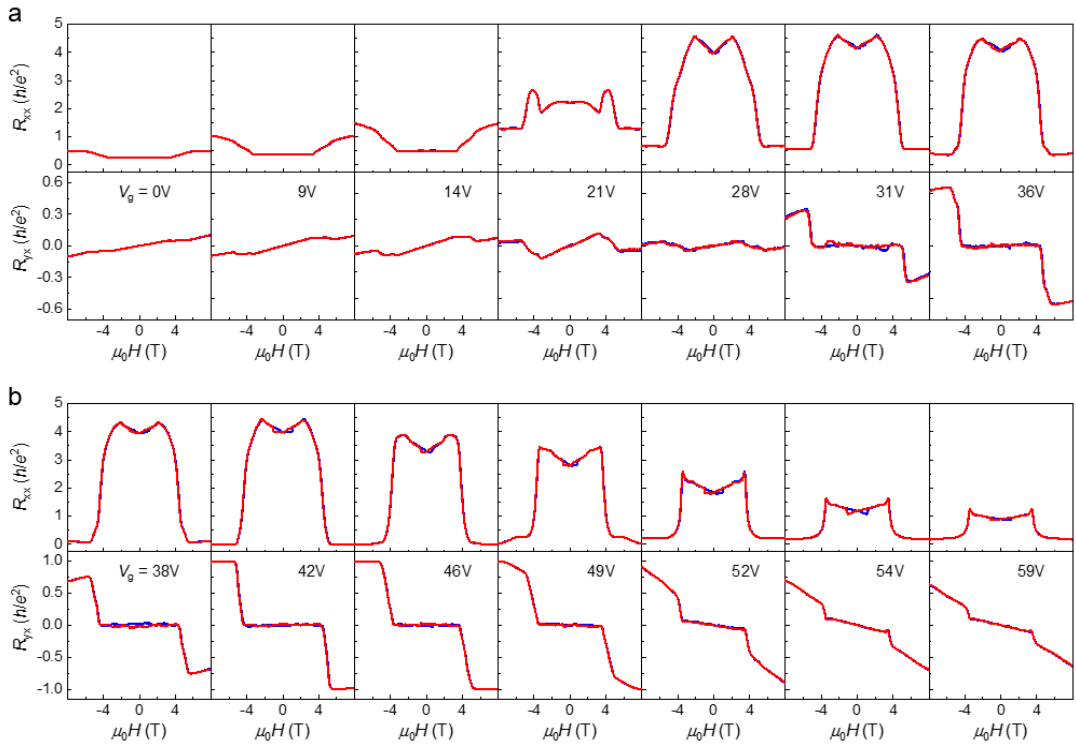

**Supplementary Fig. 4 | Magnetic field dependent $R_{xx}$ and $R_{yx}$ at 1.6 K for varied**

$V_g$s for Device #6-SL-1. Magnetic field dependent $R_{xx}$ and $R_{yx}$ measured at 2 K for 0 V ≤ $V_g$ ≤ 36 V (a), and 38 V ≤ $V_g$ ≤ 59 V (b). The best regime for the Chern insulator phase lies in $V_g$ from 42 V to 46 V.

Supplementary Fig. 5 displays the magnetic field dependent $R_{xx}$ and $R_{yx}$ at varied temperatures measured at $V_g$ = 46 V for Device #6-SL-1. Slightly quantitative difference between the data shown here and that in Supplementary Fig. 4 is because this series of data was acquired after the $V_g$ dependent measurements. Supplementary Fig. 5 clearly shows the zero-field $R_{xx}$ increases with the decrease of temperature, reaching as high as 4 $h/e^2$ at the lowest temperature 1.6 K. Accompanied by the insulating behavior of $R_{xx}$, $R_{yx}$ displays a wide zero Hall plateau at the same field range. Both behaviors are characteristics of the axion insulator phase. With the increase of magnetic field, the AFM state is driven to the FM state, and $R_{xx}$ quickly drops to nearly zero. Accordingly, $R_{yx}$ undergoes a sharp transition from the zero plateau to the $R_{yx}$ = -$h/e^2$ plateau. The quantized $R_{yx}$ and vanished $R_{xx}$ undoubtedly demonstrate that the 6-SL $MnBi_2Te_4$ device is in the Chern insulator phase for the FM state. Remarkably, the quantization of the Chern insulator phase here is highly stable against the thermal activation. Even at 8 K, |$R_{yx}$| is still quantized at 0.997 $h/e^2$ and $R_{xx}$ is as low as 0.009 $h/e^2$.

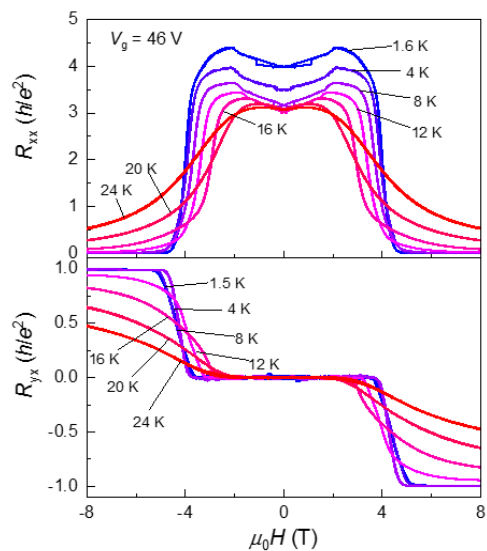

**Supplementary Fig. 5 | Magnetic field dependent $R_{xx}$ and $R_{yx}$ at $V_g$ = 46 V for**

varied temperatures for Device #6-SL-4. At $\mu_0H$ = 8 T, the Chern insulator phase persists to temperature as high as 8 K, with $|R_{yx}|$ higher than 0.997 $h/e^2$, and $R_{xx}$ as small as 0.009 $h/e^2$.

**Supplementary Note 5:**
**$V_g$ dependent $R_{xx}$ and $R_{yx}$ for Device #6-SL-1 in pulsed magnetic fields**

Supplementary Figs. 6a and 6b display the $V_g$ dependent $R_{xx}$ and $R_{yx}$ in pulsed magnetic fields for the 6-SL device. The entire data set were acquired after the measurement at Tsinghua. As mentioned above, the electrical contact is fragile against the sample transfer process, in particular after undergoing frequent thermal cycling process. One current contact is completely broken when transferred to the pulsed magnetic field facility. Thus, it is replaced by a side voltage contact during the pulsed field measurements. The layout of contact arrangement is displayed in the inset of Supplementary Fig. 6a. The current flows from contact 2 to contact 4. $R_{xx}$ and $R_{yx}$ are obtained simultaneously by measuring $V_{65}$ and $V_{35}$. Such setup only quantitatively affects the value of $R_{xx}$, but would not affect the qualitative transport behaviors.

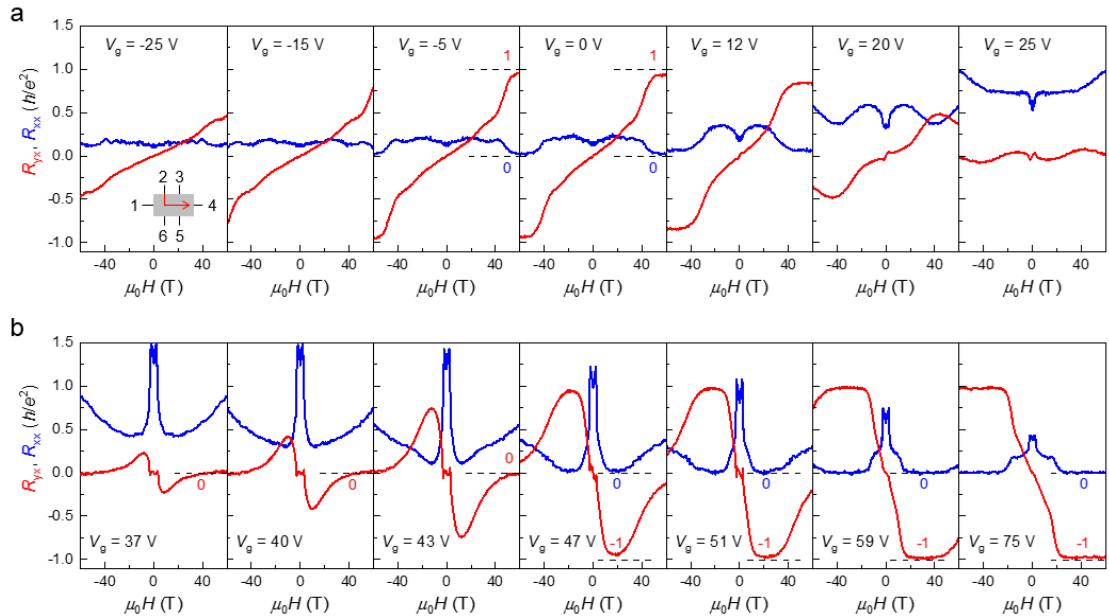

**Supplementary Fig. 6 | The $V_g$ dependent transport properties at 2 K for Device #6-SL-1 in pulsed magnetic fields.** Magnetic field dependent $R_{xx}$ (blue) and $R_{yx}$ (red) measured at -25 V ≤ $V_g$ ≤ 25 V **(a)** and 37 V ≤ $V_g$ ≤ 75 V **(b)**. The $C$ = +1 phase is most

pronounced for -5 V ≤ $V_g$ ≤ 0 V, with $R_{yx}$ = 0.94 ± 0.01 $h/e^2$ and $R_{xx}$ = 0.03 ± 0.02 $h/e^2$ at 60 T. Starting from $V_g$ = 47 V, the $C$ = -1 phase shows up when magnetic field exceeds 20 T, with $R_{yx}$ = -0.94 ± 0.01 $h/e^2$ and $R_{xx}$ = 0.02 ± 0.02 $h/e^2$ at 60 T. The best quantization is realized at $V_g$ ≥ 59 V, in which $R_{yx}$ = -0.98 ± 0.01 $h/e^2$ and $R_{xx}$ = 0 ± 0.02 $h/e^2$. The error bars are estimated from the amplitudes of the fluctuations in the signals.

Supplementary Fig. 6a shows the magnetic field dependent $R_{xx}$ (blue) and $R_{yx}$ (red) measured at $V_g$ ranging from -25 V ≤ $V_g$ ≤ 25 V. In this $V_g$ range, the transport is mainly conducted by hole-like carriers, as reflected by the overall positive slope of $R_{yx}$. Starting from $V_g$ = -5 V, pronounced QH state with $C$ = +1 appears in high magnetic fields. At 60 T, $R_{yx}$ reaches 0.94 ± 0.01 $h/e^2$, and $R_{xx}$ drops to as low as 0.03 ± 0.02 $h/e^2$. Such feature is also observed in the 7-SL thick devices, as presented in Fig. 2 in the main figures. As $V_g$ is further increased to 37 V, the $C$ = 0 phase appears.

Although Device #6-SL-1 exhibits qualitatively consistent behaviors as that of the three 7-SL devices, there are several quantitative differences. Firstly, in the 6-SL device, 60 T is not large enough for the simultaneous observation of $C$ = -1 and $C$ = 0 phases. The plateau width of the $C$ = 0 phase is much narrower than that of the 7-SL devices, and only appears in the highest magnetic field regime. Secondly, unlike the behaviors for the 7-SL devices (Fig. 2 in the main figures and Supplementary Fig. 2), where the $C$ = 0 phase becomes broader and the $C$ = -1 phase shifts towards the low-field side as holes are injected, the $C$ = 0 phase for Device #6-SL-1 remains in a narrow high field regime. Several possible reasons could account for these discrepancies. One intrinsic reason is that the gap size of the trivial quantum well bands in the 6-SL sample is larger than that of the 7-SL sample, thus a much larger magnetic field is required to completely suppress the $C$ = -1 phase. A more likely reason is that the 6-SL device was doped due the aging effect. The unavoidable exposure to air during the sample transfer process may introduce chemical doping to the sample. The enlarged $V_g$ and magnetic field for observing the $C$ = -1 phase in pulsed field measurements provide several informative evidences. The required $V_g$ and magnetic field are 42 V and 5 T for the measurement in static magnetic fields (Supplementary Fig. 4), whereas over 47 V and 20 T are required for the pulsed magnetic fields (Supplementary Fig. 6). Moreover, the level of quantization also decreases from more than 0.99 $h/e^2$ at $V_g$ = 42 V (Supplementary Fig. 4) in the static measurements to 0.94 $h/e^2$ at $V_g$ = 47 V (Supplementary Fig. 6) in the pulsed measurements, which undoubtedly indicates that the sample quality has indeed seriously degraded.

To avoid the influence of the degradation of sample quality on our main conclusion, we performed all the measurements on Device #7-SL-1 in pulsed magnetic fields. We first characterize the detailed gate voltage dependence by applying a pulse field to 12 T, and then extend the measurements up to 61.5 T for the representative gate voltages in the interesting regimes. The results presented in the main text and figures are all from the Device #7-SL-1.

**Supplementary Note 6:**
**Experimental phase diagram of the 6-SL device**

Supplementary Fig. 7 displays the experimental phase diagram for Device #6-SL-1 summarized from the values of $R_{yx}$, $dR_{yx}/dH$ and $R_{xx}$. Starting from $V_g$ = -30 V, hole-type carriers are gradually depleted with increasing $V_g$. At $V_g$ = -5 V, QH phase with $C$ = 1 appears in the high-field regime, as represented by red and white in Supplementary Figs. 7a and 7b. As $V_g$ is increased to about 30 V, the $C$ = 0 phase appears and persists to 60 T. Similar to the phase diagram of 7-SL devices, the $C$ = 0 phase occupies the largest portion of the phase diagram and is highly stable with respect to the changes of both $V_g$ and magnetic field. Further increase of $V_g$ leads to the appearance of the $C$ = -1 and $C$ = -2 phase. Compared to the $C$ = 1 phase in the hole-type regime, the area of $C$ = -1 phase is much wider. The 6-SL and 7-SL devices manifest consistent phase diagrams, indicating that the observed phenomena are universal for $MnBi_2Te_4$ in the 2D limit.

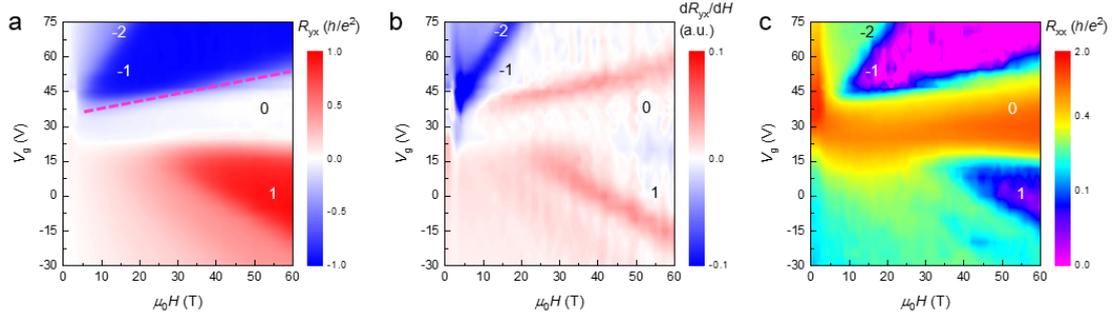

**Supplementary Fig. 7 | Experimental phase diagram for Device #6-SL-1.** The measured $R_{yx}$ (**a**), $dR_{yx}/dH$ (**b**), and $R_{xx}$ (**c**) values as functions of magnetic field and $V_g$.

**Supplementary Note 7:**
**Schematic band structure evolution for different situations of Zeeman effect**

In Supplementary Fig. 8, we compare the band structure evolution in magnetic field in the three situations of without Zeeman effect, with small and with large Zeeman effect. Without Zeeman effect, the energy gap is only determined by the cyclotron motion of electrons, which increases in magnetic field. The gap size enlarges as the formation of Landau levels, which is also demonstrated by the calculated Landau level spectrums. As shown in Supplementary Fig. 8a, for the $C = -1$ phase with $E_F$ lying in the band gap (green shadow regime), it will persists in magnetic field. And for $E_F$ lying in the valence band, increasing magnetic field will further stabilizes the $C = -1$ phase. Obviously, both are absent in our experiment. Supplementary Fig. 8b shows the case when there is a small Zeeman effect. In this case, the two blue bands move oppositely in magnetic field and the band gap progressively decreases. It is clear that for $E_F$ lying in the valence band ($V_g = 0$ V in our experiment), with the formation of $n = 0$ Landau levels, a $C = 0$ phase with a broad zero Hall plateau forms in magnetic field, as observed in our experiment. However, because the band gap does not close, the $C = -1$ Chern insulator phase for $E_F$ lying in the gap is unaffected in magnetic field (green shadow regime). As long as there is no band inversion, such $V_g$ regime for $C = -1$ phase is persistent forever. Therefore, considering Landau levels and a small Zeeman effect is also insufficient for our experiment. To fully explain our experimental data, not only Zeeman effect is indispensable, but also its magnitude is required to be sufficiently large so that a band inversion can be induced, as shown in Supplementary Fig. 8c. In this case, all the experimental observation in our work can be well explained.

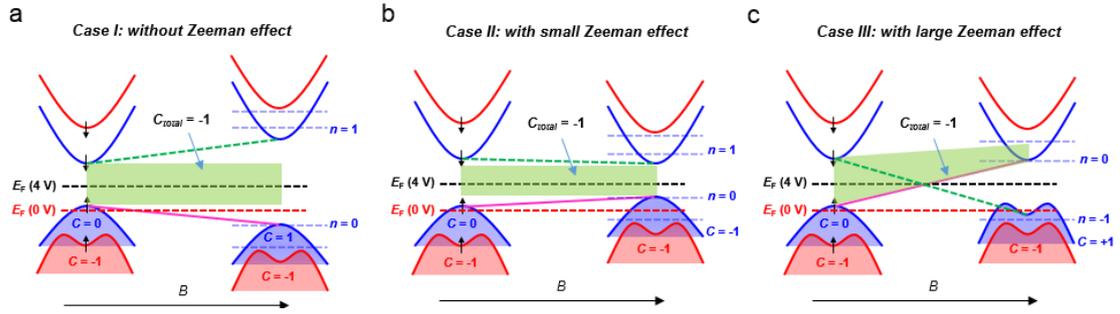

**Supplementary Fig. 8 | Schematic band structure evolution without Zeeman effect (a), with small Zeeman effect (b) and with sufficient large Zeeman effect (c).** The green shadow regime represents the $V_g$ range where the $C = -1$ Chern insulator phase persists in strong magnetic field.

**Supplementary Note 8:**
**Effective Hamiltonian and calculated Landau levels**

To understand the experimental findings, we performed first-principles electronic structure calculations and studied the magnetic/topological properties by the effective Hamiltonian method. In zero magnetic field, the bulk of MnBi$_2$Te$_4$ is an AFM TI, and the (111) films have topological surface states gapped by magnetism. An interesting property of MnBi$_2$Te$_4$ in AFM state is that the electronic coupling between neighboring SLs is largely suppressed by the parity-time (*PT*) symmetry[3]. By applying strong out-of-plane magnetic fields, an AFM to FM phase transition happens, which results in *PT* symmetry breaking and thus significantly enhances the interlayer coupling. This could drive a topological phase transition from AFM TI to Weyl semimetal, according to theoretical calculations[4,5]. Theoretically, the low-energy physics of FM MnBi$_2$Te$_4$ films are described by a four-band model based on the four quantum well states near the $E_F$. Figure 3c in the main text displays the theoretical band structure for the 7-SL MnBi$_2$Te$_4$ film with FM order. Remarkably, the second highest valence band displays an obvious M-shape feature near Γ, which is a signature of topological band inversion. Topological edge-state calculations reveal that the band inversion indeed happens between this valence band and a conduction band, as shown in Supplementary Fig. 9. We also computed the topological invariant and got $C = -1$, in good agreement with the experiment. Bands near the Fermi level are mainly contributed by the $p_z$ orbitals of Te atoms, which have small orbital angular momentum. For out-of-plane polarized $Mn^{2+}$ moments, the $z$ components of spin angular momentum $s_z$ of the first/second conduction band minimum and valence band maximum (named $CBM_1/CBM_2$ and $VBM_1/VBM_2$) are negative and positive, respectively, corresponding to total angular momentum $J_z = -1/2$ and $+1/2$ (Fig. 3d in main text) as found by first-principles calculations.

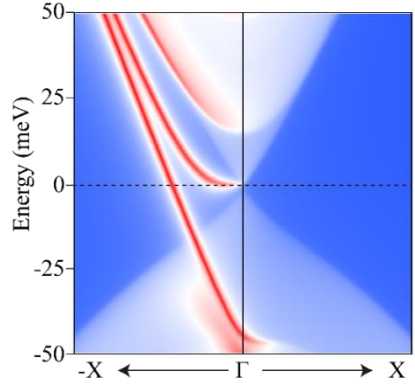

**Supplementary Fig. 9 | Edge states along (100) direction in the 7-SL FM MnBi$_2$Te$_4$ film.**

Using the bases of $|CBM_1\rangle$, $|VBM_1\rangle$, $|VBM_2\rangle$, and $|CBM_2\rangle$, which have $J_z = -1/2$, $+1/2$, $+1/2$ and $-1/2$, respectively, we constructed a low-energy effective Hamiltonian by the $\boldsymbol{k \cdot p}$ approach[6]:

$$H(k) = \begin{bmatrix} h_+(k) & 0 \\ 0 & h_-(k) \end{bmatrix},$$

where $h_\pm(k) = \epsilon_\pm(k)\sigma_0 + \boldsymbol{d}_\pm \cdot \boldsymbol{\sigma}$, $\sigma_{x,y,z}$ is the Pauli matrix, $\epsilon_\pm(k) = \epsilon_0^\pm + D_\pm k^2$, $k^2 = k_x^2 + k_y^2$, $\boldsymbol{d}_\pm = (A_\pm k_x, \pm A_\pm k_y, M_0^\pm - M_2^\pm k^2)$, and $M_2^\pm < 0$ is assumed. Importantly, $h_\pm(k)$ gives Chern number $C_\pm = 0$ when $M_0^\pm > 0$ and $C_\pm = \pm 1$ when $M_0^\pm < 0$ (ref. [7,8]). By fitting bands from *ab initio* calculations, we obtained a positive $M_0^+$ and a negative $M_0^-$, giving $C_+ = 0$ and $C_- = -1$. The total Chern number is $C = C_+ + C_- = -1$. As the two bands given by $h_-(k)$ are relatively far away from the $E_F$, magnetic responses of the system are mainly determined by $h_+(k)$. The fitting parameters of $h_+(k)$ are $\epsilon_0^+ = 0$ eV, $D_+ = 16.0$ eV·Å$^2$, $A_+ = 1.5$ eV·Å, $M_0^+ = 1.5 \times 10^{-3}$ eV, and $M_2^+ = -16.8$ eV·Å$^2$.

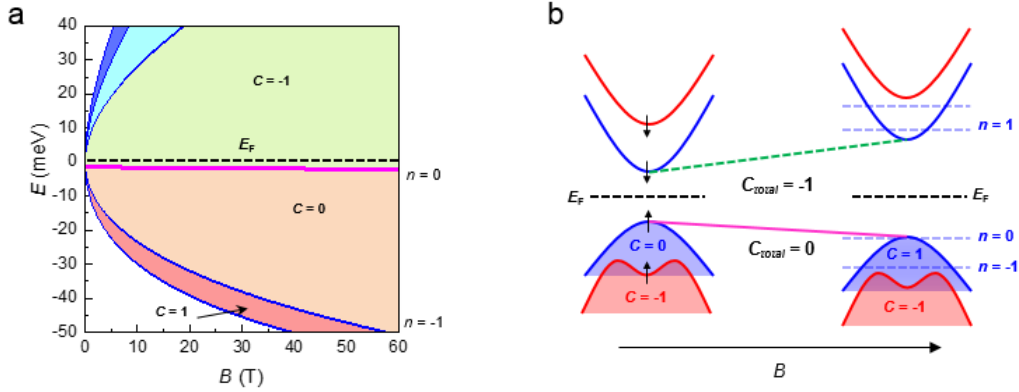

**Supplementary Fig. 10 | Calculated Landau level spectrums and Chern numbers in magnetic field without Zeeman effect (a) and schematic band structure evolution (b).**

Then we studied the influence of external magnetic field $\boldsymbol{B} = (0, 0, B)$ based on $h_+(k)$. We chose the Landau gauge $\boldsymbol{A} = (-By, 0, 0)$ and included the orbital effect by replacing the electron momentum $\boldsymbol{p} \to \boldsymbol{p} + e\boldsymbol{A}$, where $e$ is the elementary charge ($e > 0$). The Zeeman effect is described by $H_{\text{Zeeman}} = -\frac{1}{2} g \mu_B B \sigma_z$, where $g$ is the g-factor ($g \approx 2$ for free electrons) and $\mu_B$ is the Bohr magneton. The energy spectra of Landau levels for $h_+(k)$ can be analytically calculated[9]:

$$E_n = \begin{cases} \epsilon_0^+ + \frac{eB}{\hbar}(2nD_+ + M_2^+) \pm \sqrt{2n\frac{eB}{\hbar}A_+^2 + \left(M_0^+ - \frac{1}{2}g\mu_B B - \frac{eB}{\hbar}(D_+ + 2nM_2^+)\right)^2}, \\ \qquad\qquad\qquad n = 1, 2, 3\ldots \\ \epsilon_0^+ - M_0^+ + \frac{1}{2}g\mu_B B + \frac{eB}{\hbar}(D_+ + M_2^+), \ n = 0. \end{cases}$$

If neglecting the Zeeman effect by selecting $g = 0$, one would always get the $C = -1$ phase ($C_+ = 0$ and $C_- = -1$) at the charge neutral regime, as displayed in Supplementary Fig. 10a, even in a large magnetic field, which obviously contradicts with the experiment. Supplementary Fig. 10b shows the schematic illustration of the band structure evolution in magnetic field without Zeeman effect. It clearly shows that as long as the $E_F$ (black dashed line) is in the band gap (corresponding to the $V_g$ range in Fig. 2a), it will remain forever in the $C = -1$ phase in high magnetic field. It is thus concluded that the Zeeman-effect-induced band inversion plays a crucial role in driving the $C = -$ to $C = 0$ phase transition. Notably, it is difficult to evaluate the exact $g$-factor by first-principles calculations. By fitting the theoretical phase diagram with experiment and referring to previous experimental results in conventional TIs[10-12], $g = 10$ is selected in our calculations, which is on the same order of magnitude as $Bi_2Te_3$ (ref. [13]).

## Supplementary Note 9:
## Reproducible nonlocal transport data for Device #7-SL-3

In Supplementary Fig. 11, we display the results of four-probe and two nonlocal measurements with different configurations in a third 7-SL device. As the increase of magnetic field, $R_{yx}$ decreases from $-h/e^2$ towards 0, accompanied by the saturation of $R_{xx}$ towards 0.5 $h/e^2$, highly consistent with the data of Device 7-SL-1 shown in Fig. 2 in the main figures. Noting that the transition from the $C = -1$ to $C = 0$ phase in this device is not as sharp as that in Device 7-SL-1. This broad transition accidently allows us to observe the signature of quantized $R_{xx}$ in nonlocal measurements in a broader magnetic field range, due to the weak scattering between the counter-propagating edge states at the onset field regime. Supplementary Figs. 10b and 10c display the nonlocal measurements results using the same configurations as Fig. 4b in the main figures. Clearly, broad $R_{xx}$ plateaus with different quantized values are observed. These results unambiguously demonstrate the reproducibility of the nonlocal data.

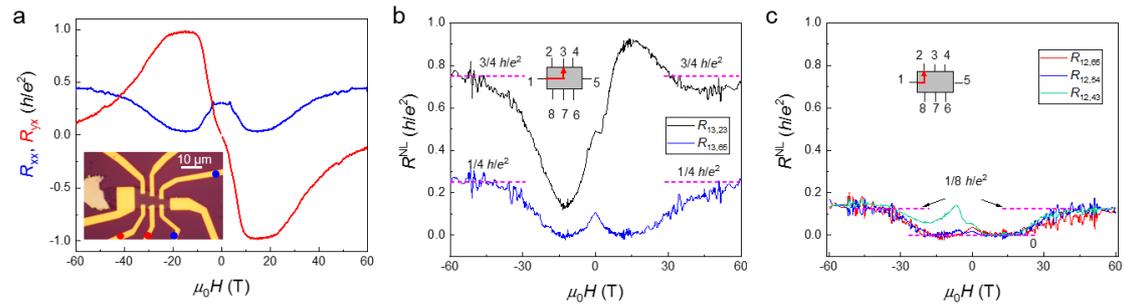

**Supplementary Fig. 11 | Nonlocal transport data for Device #7-SL-3. a,** Magnetic field dependence of $R_{xx}$ and $R_{yx}$ in four-probe measurement. The inset shows the optical image of this device. The blue and red dots denote the selected contacts for $R_{yx}$ and $R_{xx}$ measurements. **b,** Nonlocal measurements with current flowing between electrodes 1 and 3. Depending on the position of the voltage probes, the resistance is nearly quantized at 1/4 and 3/4 $h/e^2$. **c,** Nonlocal measurements with current flowing between electrodes 1 and 2. All the $R_{xx}$ converge to 1/8 $h/e^2$. The expected quantized values of $R_{xx}$ are denoted by the magenta broken lines.